\documentclass[11pt,twoside]{article} 
\usepackage{asp2004}
\usepackage{epsf}
\usepackage{psfig}
\usepackage{lscape} 

\begin{document} 
\title{Spectral Analysis of Central Stars of PNe Interacting with the Interstellar Medium}
\author{T. Rauch}
\affil{Dr.-Remeis-Sternwarte, Sternwartstra\ss e 7, 96049 Bamberg, Germany}
\affil{Institut f\"ur Astronomie und Astrophysik, Sand 1, 72076 T\"ubingen, Germany}
\author{F. Kerber} 
\affil{Space Telescope -- European Coordinating Facility,
       European Southern Observatory,
       Karl-Schwarzschild-Stra\ss e 2, 85748 Garching, Germany}
\begin{abstract} 
Planetary Nebulae (PNe) are the result of heavy mass loss of the
asymptotic giant branch (AGB) stars. They are understood in terms
of Kwok's (1978) interacting-winds model as the product of the mass-loss
history on the AGB and the central star (CS) evolution. Since the CS are close 
to the end of nuclear burning and at their hottest stage of evolution then, precise modeling
of these pre-white dwarfs is a prerequisite in order to calculate reliable
ionizing fluxes which are crucial input for the presently available
3D photoionization codes. In the framework of a 
systematic study of PNe which show evidence for an interaction with the 
ISM, 
we present a NLTE analysis of their CS.
\end{abstract}

\section{Introduction}
Recent developments in both, observational and deprojection techniques, spectral
analysis, and numerical methods facilitate to closely examine and model
PNe and their CS consistently.
Spectral analysis of CSPN by means of modern NLTE model atmosphere techniques
provides information about photospheric parameters. 
In comparison with evolutionary calculations,
evolutionary status, distances, masses, and luminosities can be determined.
The spectroscopic distances enable us to deal with precise linear diameters
of the PNe. The stellar model fluxes are an important input as realistic ionizing 
spectra in nowadays available 3D photoionization codes (e.g\@. Ercolano et al\@. 2003)
for analyses of the PNe. The consistent modeling of PN and CS allows to determine
the yields of nuclear processed material which goes back to the ISM via PNe and
thus, to investigate the chemical evolution of our Galaxy.

In the last decade, PNe which show evidence for an interaction with the ISM
have been found. This process is an indicator for the evolution of
their CS which are at their hottest stage of evolution close to the end of 
nuclear burning where gravitational effects become dominant (i.e\@. they display 
directly the formation of white dwarfs).
The highly evolved CS are no longer dominating the processes in the PNe
(Kerber \& Rauch 2001); the nebulae display brightness asymmetries that 
reflect the degree of the interaction process.
These complex objects are crucial tests for our models as well as for
evolutionary theory. 

Once regarded a curiosity, this complex PN -- ISM 
interaction process appears to be a common phenomenon which allows 
to determine also properties of the ambient ISM. Recently, we have 
presented determinations of proper motions and Galactic orbits
(Kerber et al\@. 2004a, 2004b). These unveil thin and thick disk populations
and cast light on interaction with the ISM. First
hydrodynamical test calculations (M\"uller et al\@. 2004) for these 
objects were performed with the aim of a more quantitative
understanding of processes in PNe in decay.
One of the pre-requisites for reliable modeling are spectral analyses
of the CS. We present here preliminary results from a study
which is based on line-blanketed models including H and He. 

\section{Observations}

In July 1999, we performed medium-resolution spectroscopy of nine CS with
the TWIN spectrograph attached to the 3.5m telescope at Calar Alto, Spain.
The CS of A\,21 was observed in January 1999 with EFOSC\,1 at the 3.6m
telescope of ESO, La Silla.

\begin{table}[t]
\caption{Exposure times of our programme stars}
\smallskip
\begin{center}
{\small
\begin{tabular}{llcc}
\tableline
\noalign{\smallskip}
\noalign{\smallskip}
name & PN\,G & $m_\mathrm{v}$ & exposure time \\
     &       &                & [sec] \\
\noalign{\smallskip}
\tableline
\noalign{\smallskip}
A\,21            & 205.1+14.2   & 16 & 7200 \\
A\,52            & 050.4+05.2   & 18 & 3600 \\
A\,75            & 101.8+08.7   & 18 & 7200 \\
DeHt\,5          & 228.2$-$22.1 & 15 & 6000 \\
EGB\,1           & 124.0+10.4   & 17 & 6300 \\
NGC\,6781        & 041.8$-$02.9 & 17 & 2400 \\
NGC\,6842        & 065.9+00.5   & 16 & 2400 \\
RX\,J2117.1+3412 & 080.3$-$10.4 & 13 & 1200 \\
Sn\,1            & 013.3+32.7   & 15 & 1200 \\
WeSb\,5          & 058.6$-$05.5 & 17 & 3600 \\
\noalign{\smallskip}
\tableline
\end{tabular}
}
\end{center}
\end{table}

\section{ NLTE Model Atmospheres}

For the classification and preliminary analysis of hot compact stars, we have set up a
new grid of H+He models within
$T_\mathrm{eff}$ = 50 - 190\,kK in 10\,kK steps, 
$\log g$         =  5 -   9     in 0.5 steps (cgs), 
H/He             =  0 -   1     in 0.1 steps by mass. 
The plane-parallel, static models are calculated with TMAP, the T\"ubingen NLTE Model Atmosphere Package 
(Werner et al\@. 2003, Rauch \& Deetjen 2003). 
The grid which we used for this analysis and some other grids of NLTE model atmosphere fluxes with different chemical
composition will be available on the WWW 
({\tt http://astro.uni-tuebingen.de/\raisebox{1.5mm}{{\tiny$\sim$}}\hspace{0.3mm}rauch}).

\begin{figure}[ht]
\plotone{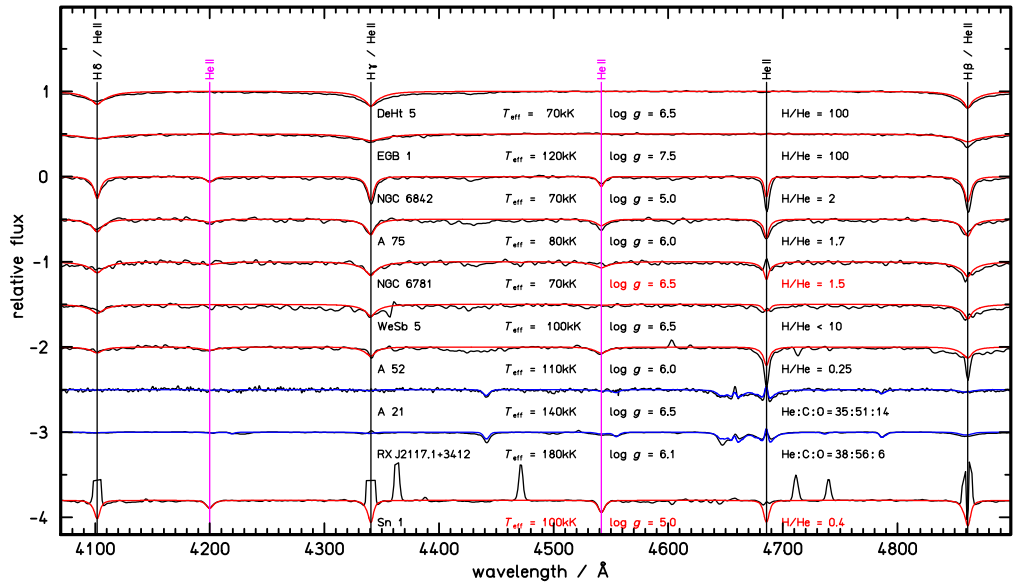}
\caption{Comparison of synthetic spectra with the observation.}
\label{fit}
\end{figure}

We aim to arrive at a maximum error in our preliminary spectral analysis of about 1\,dex in $T_\mathrm{eff}$, 
0.3\,dex in $\log g$, and 0.3\,dex in the H/He abundance ratio. We employed a $\chi^2$ fit procedure in order
to judge our fit-by-eye procedure. In those cases where we have almost pure H or He atmospheres, the deviations
were smaller than 10\,kK in $T_\mathrm{eff}$ and 0.3\,dex in $\log g$. However, since the S/N of our spectra was
not very high and we have partly relics from nebular emission which even survived a careful data reduction, we
present here our fit-by-eye results. These are summarized in Table~\ref{php} and Fig.~\ref{fit}.

\begin{table}[t]
\caption{Parameters of our programme stars. Detailed analyses
of the CSPN of DeHt\,5 and EGB\,1 have been presented by 
Barstow et al\@. 
(2003, $T_\mathrm{eff}\hspace{-0.5mm} =\hspace{-0.5mm} 58\,582\,\mathrm{K}$ and $\log g\hspace{-0.5mm} =\hspace{-0.5mm} 7.05$) 
and Napiwotzki 
(1999, $T_\mathrm{eff}\hspace{-0.5mm} =\hspace{-0.5mm}    147\,\mathrm{kK}$ and $\log g\hspace{-0.5mm} =\hspace{-0.5mm} 7.34$),
respectively.
The CSPN of A\,21 and RX\,J2117.1+3412 have been analyzed by Rauch \& Werner (1995) and 
Rauch \& Werner (1997), respectively.
Stanghellini et al\@. (2002) presented Zanstra temperatures for the CSPN of 
NGC\,6842 ($97\,\mathrm{kK}$),
A\,75     ($< 290\,\mathrm{kK}$),
and
NGC\,6781 ($105\,\mathrm{kK}$).
}
\smallskip
\begin{center}
{\small
\label{php}
\begin{tabular}{lccc}
\tableline
\noalign{\smallskip}
name & $T_\mathrm{eff}$ & $\log g$ & H/He   \\
     & [kK]             & [cgs]    & [mass] \\
\noalign{\smallskip}
\hline
\noalign{\smallskip}
DeHt\,5          &  70 & 7.5 & $>$100 \\
EGB\,1           & 120 & 7.5 & $>$100 \\
\tableline           
\noalign{\smallskip}
NGC\,6842        &  70 & 5.0 &   2    \\
A\,75            &  80 & 6.0 &   1.7  \\
NGC\,6781        &  70 & 6.5 &   1.5  \\
WeSb\,5          & 100 & 6.5 & $<10$   \\
Sn\,1            & 100 & 5.0 &   0.4  \\
A\,52            & 110 & 6.0 &   0.25 \\
\tableline           
\noalign{\smallskip}
A\,21            & 140 & 7.5 & \multicolumn{1}{l}{He:C:O=35:51:14} \\
RX\,J2117.1+3412 & 180 & 6.1 & \multicolumn{1}{l}{He:C:O=38:56:6}  \\
\tableline
\end{tabular}
}
\end{center}
\end{table}

\section{Results}

Within our sample of ten CS of PNe which show interaction with the ISM, our
spectral analysis includes two hydrogen-rich DA (pre-) white dwarfs (DeHt\,5,
EGB\,1), two hydrogen-deficient PG\,1159 stars (A\,21 and RX\,J2117.1+3412), and six
CS with intermediate H/He ratios (from 0.25 to 10 by mass).

Fine tuning of the parameters with models which consider metal opacities
in the next part of this analysis will enable us to determine e.g\@. their spectroscopic
distances reliably. These are necessary to calculate the linear dimensions of the PNe for hydrodynamical
modeling. 

\acknowledgements{This research was supported by the DFG under grants RA\,733/3-1 and RA\,733/14-1,
                  and by the DARA/DLR under grants 50\,OR\,9705\,5 and 50\,OR\,0201.}

\clearpage

\end{document}